\documentclass[sigconf]{acmart}
\usepackage{xspace}
\usepackage{dsfont}
\usepackage{setspace}
\usepackage{enumitem}
\usepackage{setspace}
\usepackage{algorithm}
\usepackage{algpseudocode}
\usepackage{xcolor}
\usepackage{tikz}
\usetikzlibrary{arrows.meta,calc}
\graphicspath{{figures/}}

\AtBeginDocument{%
  }

\setcopyright{acmlicensed}
\copyrightyear{2018}
\acmYear{2018}
\acmDOI{XXXXXXX.XXXXXXX}
\acmConference[Conference acronym 'XX]{XXX}{June 03--05,
  2018}{Woodstock, NY}
\acmISBN{978-1-4503-XXXX-X/2018/06}

\begin{document}

\title{PinGER: Token-Efficient Generative Retrieval with Semantic IDs}

\newcommand{\pinaff}{\affiliation{\institution{Pinterest}\city{Palo Alto}\state{CA}\country{USA}}}
\author{Xinyi Li}
\email{xinyi@pinterest.com}
\pinaff
\author{Sujan Reddy Ale}
\email{sreddyale@pinterest.com}
\pinaff
\author{Yi-Ping Hsu}
\email{yhsu@pinterest.com}
\pinaff
\author{Zelun Wang}
\email{zelunwang@pinterest.com}
\pinaff
\author{Yilin Chen}
\email{yilinchen@pinterest.com}
\pinaff
\author{Hanyu Li}
\email{hanyuharryli@pinterest.com}
\pinaff
\author{Hanlin Lu}
\email{hanlinlu@pinterest.com}
\pinaff
\author{Arkin Dharawat}
\email{adharawat@pinterest.com}
\pinaff
\author{Lulu Yao}
\email{yyao@pinterest.com}
\pinaff
\author{Yaxin Li}
\email{yaxinli@pinterest.com}
\pinaff
\author{Bella Huang}
\email{shuang@pinterest.com}
\pinaff
\author{Xiangyi Chen}
\email{xiangyichen@pinterest.com}
\pinaff
\author{Jaewon Yang}
\email{jaewonyang@pinterest.com}
\pinaff
\author{Jiajing Xu}
\email{jiajing@pinterest.com}
\pinaff
\author{Charles Rosenberg}
\email{crosenberg@pinterest.com}
\pinaff
\renewcommand{\shortauthors}{Li et al.}

\begin{abstract}
Generative retrieval replaces nearest-neighbor search with token-by-token decoding of an item’s semantic ID (SID). But existing SID-based methods primarily use $L$ levels of tokens sequentially to represent each item in the input history. As a result, a fixed context window holds $L$ times fewer items than it would with one embedding per item.
We present PinGER, a generative retriever deployed at Pinterest that retains SID outputs while making input histories compact, built on three ideas. (1) An asymmetric
attention mask over interleaved SID and dense tokens lets later items attend
only to the dense ones, so the model is supervised at every SID position in
training yet serves one token per item while computing the same function. (2)
Generation is controllable through goal-conditioned outputs, reinforcement
learning, and a prefix-diversity penalty, which together steer retrieval toward
surface-specific goals and diverse candidates. (3) A serving stack built for
efficient wide-beam generation is paired with decoding constrained to valid
identifiers and candidate resolution from a host-local index. Deployed in
Related Pins, PinGER lifts saves of fresh Pins by $3.60\%$ and overall
saves by $0.62\%$, while reducing the surface's end-to-end latency by
$0.7\%$ and eliminating the cost of the nearest-neighbor search it
replaces.

\end{abstract}



\maketitle

\section{Introduction}\label{sec:intro}

\begin{figure*}[t]
  \centering
  \includegraphics[width=\textwidth]{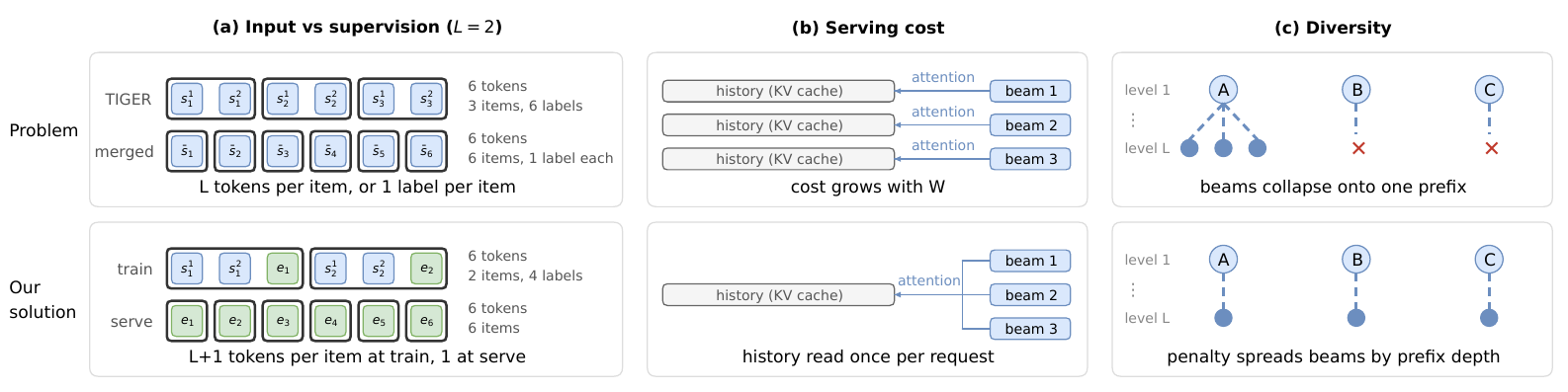}
  \caption{Three obstacles to efficient SID retrieval (top) and our answers
  (bottom). $s$: SID token, $e$: dense token.}
  \label{fig:overview}
\end{figure*}

Retrieval is the first stage of an industrial recommender: from a corpus of
order $10^{9}$ items, it selects the candidates a downstream ranker will score,
under a tight latency budget. The dominant approach is two-tower retrieval with
approximate nearest neighbor (ANN) search~\cite{twotower}: user and item towers
embed both sides in a shared space and retrieval returns the items nearest the
user vector, so the two interact only through a final dot product. Since sequential recommenders scale like language models~\cite{hstu},
generative retrieval (GR) instead generates the next item as a language model
generates the next token~\cite{tiger}.

A GR model can represent items in one of two forms.
\emph{Embedding GR}, such as PinRec~\cite{pinrec}, reads and predicts one dense
embedding per item, so a fixed context covers a long history, but it still
retrieves through an ANN index. \emph{SID GR}, introduced by TIGER~\cite{tiger},
quantizes each item into a semantic ID (SID) of $L$ discrete codes and
generates it with beam search, so the output space can be decoded directly
into items and steered during generation. As a result, no ANN index sits on the
request path, and a new item is retrievable as soon as it has an SID. Its
inputs are costly, however, since every history item takes $L$ tokens, even
though history matters: Recall grows by $10\%$ as the training context grows from
$1{,}024$ to $6{,}144$ tokens (Section~\ref{sec:seq-length}). We ask whether one
model can have both, reading history at one token per item while keeping SID
outputs, their supervision, and the diversity their output space allows.

Three obstacles stand in the way (Figure~\ref{fig:overview}). The first is that input and supervision pull
in opposite directions. In a causal model, every token the model is trained on
is also input for every later token. SID tokens give rich supervision, one
label per code, but cost $L$ input tokens per item, and compressing them removes the
token-level supervision. Naive fixes pick one side: a separate dense
encoder and SID decoder loses direct attention to distant history and gets one
label per forward pass, while merging each item's SIDs into one token keeps
the input short but trains on one label per item instead of $L$. The second
is beam decoding cost. A KV cache avoids recomputing the
history~\cite{pagedattention}, but every beam
still attends over all of it, so cost grows with beam width, and some
beams lead to no matching item. The third is diversity.
Because beam scores accumulate over levels, the beams collapse onto a few
prefixes ($11.9$ distinct level-$1$ codes among $100$ beams;
Table~\ref{tab:div-sweep}), and temperature sampling, the usual remedy, lowers
spread further and produces more SIDs that do not exist in the corpus.

Prior work leaves these obstacles open. On input and supervision, every prior
design pays on one side. Deployed SID retrievers keep multi-token SIDs in the
input and pay at serving~\cite{tiger,plum,onerec,onerecv2,snaplgr}, as do
methods that restructure SIDs to cut decoding steps~\cite{singlelevel,rpg,setrec};
COBRA~\cite{cobra} and GPR~\cite{gpr} add a dense token or a second decoder on
top and pay even more. SPARC~\cite{sparc} and ACERec~\cite{acerec} take the
other side: they merge each item's SIDs into one input token, so they train on
one label per item instead of $L$. On serving cost,
speculative decoding~\cite{atspeed,nezha} and serving co-design~\cite{grace}
accelerate GR but keep full-model wide-beam decoding, and shared-prefix
attention~\cite{hydragen} has not been applied to SID beam search. On
diversity, no GR work diversifies beam search, so the usual knob remains
temperature. Finally, embedding GR and SID GR have never been compared head
to head online; the only direct comparison, LIGER~\cite{liger}, is offline.

We address all three with PinGER (Pinterest GEnerative Recommender;
Figure~\ref{fig:overview}), which interleaves the two item forms in one causal
backbone: in training, each history item contributes its $L$ SID tokens
followed by one dense token, and every SID token is a prediction target. An
asymmetric attention mask lets SID tokens read the dense history and the
shallower codes of their own item, while dense tokens read only earlier dense
tokens. No later position ever reads a past SID token, so serving drops them
and reads the history at one dense token per item, computing exactly the same
function (Section~\ref{sec:mask}). In training, sparsity-aware attention kernels compute only the entries the mask permits, so a history costs less attention than in TIGER while supplying the same supervision (Section~\ref{sec:mask-kernel}). The model is thus trained with the full SID supervision of TIGER and served at the input cost of embedding GR. For
diversity, a hierarchical prefix penalty selects beams one at a time, charging
each candidate by the depth of the prefix it shares with beams already chosen
(Section~\ref{sec:diversity}). For serving, cascade attention reads the shared
history once per request rather than once per beam, constrained decoding keeps
every beam on an SID that exists, and a host-local index resolves SIDs to
items without a network call (Section~\ref{sec:infra}).

We deploy PinGER at Pinterest and evaluate it offline and online. Offline,
it matches TIGER's Recall@100 while serving $6\times$ fewer tokens, and at
equal serving tokens it exceeds TIGER by $9.5\%$ and the production
embedding retriever, PinRec, by $3.9\times$
(Section~\ref{sec:offline}). At the deployed penalty, the beams span $7.9\times$ more distinct level-$1$
prefixes than under greedy decoding (Section~\ref{sec:div-results}). End to end, serving is 18 to
31\,ms faster at the median than PinRec at matched load
(Section~\ref{sec:serving-numbers}). Online, replacing PinRec with PinGER on
Related Pins raises saves by $0.62\%$ and saves of fresh Pins by
$3.60\%$ (Section~\ref{sec:online}). A further analysis shows that PinGER surfaces
previously undervalued Pins, breaking the engagement feedback loop of
recommender systems.

Our contributions are:
\begin{itemize}[leftmargin=1.2em,itemsep=1pt,topsep=2pt]
  \item \textbf{Decoupling input from supervision.} The first role-dependent
    attention mask for GR: SID tokens are supervised in training but read by
    no later position, so they cost nothing at serving. Unlike TIGER, which
    pays $L$ tokens per item, or merged-token models, which train on one label
    per item, PinGER matches TIGER at $1/L$ the serving tokens.
    Role-dependent masks exist outside GR~\cite{unilm,xlnet}.
  \item \textbf{Hierarchy-aware beam diversity.} The first diversified beam
    search for GR: a penalty that grows with shared-prefix depth, exact, with
    $W$ additions per step in one fused kernel per level.
  \item \textbf{Cascade decode for SID retrieval.} The first use of
    shared-prefix attention for SID beam decoding, where reparenting a beam
    reduces to moving an $L$-token suffix: $10.9\times$ faster than
    paged attention at $W{=}30$ and $32\times$ at $W{=}400$, with
    $12.6\times$ less KV memory.
  \item \textbf{Online evidence.} The first head-to-head online A/B between
    embedding GR and SID GR, deployed on Related Pins at Pinterest.
\end{itemize}

\section{Related Work}\label{sec:related}

\paragraph{Generative retrieval with semantic IDs and embeddings.}
TIGER~\cite{tiger} introduced generative retrieval for recommendation:
items are quantized into hierarchical SIDs with an RQ-VAE, and a sequence
model generates the next item's SID with beam search. The paradigm has
been deployed at YouTube~\cite{plum}, Kuaishou~\cite{onerec,onerecv2} and
Snapchat~\cite{snaplgr}; all keep multi-token SIDs on both the input and
the output side, paying $L$ tokens of context per history item and $L$
decoding levels per request. Restructuring the SIDs, with a single-level
codebook~\cite{singlelevel}, parallel generation~\cite{rpg} or
order-agnostic sets~\cite{setrec}, cuts decoding steps but leaves SIDs as
the input representation. Embedding GR such as PinRec~\cite{pinrec}
instead reads and predicts one dense embedding per item and retrieves by
nearest-neighbor search, which puts an index on the request path.
Hybrids pick one side of the
trade-off: COBRA~\cite{cobra} interleaves SIDs with a dense vector and
preserves SID supervision, but pays the extra positions at serving too;
SPARC~\cite{sparc} and ACERec~\cite{acerec} merge each item's SIDs into
one input token and train on one label per item; GPR~\cite{gpr} splits
intent and SID generation across two large decoders. We keep COBRA's
interleaving in training only, and serve on the dense tokens alone
(Section~\ref{sec:mask}). Embedding GR and SID GR have been compared
offline by LIGER~\cite{liger} and argued over~\cite{egr,caseagainst}, but
never head to head in production; we report that comparison.

\paragraph{Efficient beam decoding.}
Wide-beam SID decoding is the serving bottleneck, and prior work attacks it
from two sides. Speculative decoding with draft models or
heads~\cite{atspeed,nezha} and
serving co-design~\cite{grace} accelerate generative recommenders but
keep every beam attending over the full history. Paged KV
caches~\cite{pagedattention} and shared-prefix attention~\cite{hydragen}
share the stored history across beams, yet still read it once per beam;
we fold the beam into the query dimension so the history is read once per
request (Section~\ref{sec:paged-cache}). Decoding must also stay on SIDs
the corpus contains: we adopt the vectorized trie of
STATIC~\cite{su2026vectorizing} to constrain each level, and resolve SIDs
to items from a host-local index (Section~\ref{sec:constrained}).

\paragraph{Diverse beam search.}
In language generation, diverse beam search penalizes beams that choose
the same token at the same step~\cite{li2016diverse,vijayakumar2018diverse},
a flat penalty with no notion of hierarchy. SIDs are hierarchical, so we
charge each candidate by the depth of the prefix it shares with beams
already chosen (Section~\ref{sec:diversity}). We know of no generative
retrieval work that diversifies beam search; the usual knob is
temperature, which we show lowers both spread and validity.

\section{Method}\label{sec:method}
\subsection{Background and problem setup}
\label{sec:problem}

We study the retrieval stage of a sequential recommender, following the setup of
PinRec~\cite{pinrec} and UniPinRec~\cite{unipinrec}.
A user $u$ interacts with items drawn from a corpus $\mathcal{I}$ of order
$10^{9}$ Pins, and their history at time $t$ is the chronological sequence
\begin{equation}
  \label{eq:history}
  H(u,t)=\bigl\langle (i_1,a_1,s_1,t_1), \ldots, (i_m,a_m,s_m,t_m) \bigr\rangle,
  \quad t_1\le\cdots\le t_m\le t,
\end{equation}
in which item $i_j\in\mathcal{I}$ was acted on with action $a_j\in\mathcal{A}$
on surface $s_j\in\mathcal{S}$ at time $t_j$.
Writing $\mathcal{H}$ for the space of such histories, retrieval returns an
ordered candidate list $R(u,t)\subset\mathcal{I}$ for a downstream ranker.

\paragraph{Item representation.}
We represent each item $i$ both as a dense embedding
$\mathbf{e}_i\in\mathbb{R}^{d}$ and as an SID
$c=(c_1,\ldots,c_L)$, assigned by a tokenizer
$\phi:\mathcal{I}\rightarrow\{1,\ldots,K\}^{L}$ as $L$ residual codes over
codebooks of size $K$~\cite{tiger}, where $c_{1:d}$ denotes its depth-$d$
prefix. Retrieval is autoregressive generation of the next item's identifier,
\begin{equation}
  \label{eq:model}
  f_\theta:\ \mathcal{H}\rightarrow\Delta\bigl(\{1,\ldots,K\}^{L}\bigr),
  \qquad
  p_\theta(c\mid H)=\prod_{\ell=1}^{L} p_\theta\bigl(c_\ell \mid c_{1:\ell-1},\,H\bigr),
\end{equation}
decoded by beam search of width $W$ into a set of SIDs that are then
resolved to items.

\paragraph{Section roadmap.}
Section~\ref{sec:semantic-ids} describes the tokenizer $\phi$,
Section~\ref{sec:mask} the sequence layout, the attention mask over it, and the
training objective, and Section~\ref{sec:features} the input features and the
conditioning that steers generation.
Sections~\ref{sec:diversity} and~\ref{sec:grpo} then cover decode-time and
training-time steering; resolving generated identifiers back to items is a
serving concern and is covered in Section~\ref{sec:infra}.

\subsection{Semantic IDs}
\label{sec:semantic-ids}

SIDs are encoded from each Pin's PinCLIP~\cite{pinclip} \emph{fusion} embedding
$\mathbf{x}\in\mathbb{R}^{d}$, a content-only vector of the Pin's image and
text. An RQ-VAE~\cite{lee2022residual} quantizes $\mathbf{x}$ into an SID
$c=(c_{1},\ldots,c_{L})$ with $L{=}5$ codebooks of size $K{=}2048$, selecting
one code vector $\mathbf{q}^{(\ell)}_{c_{\ell}}$ per level. Because content alone misses the Pin--Pin
similarities induced by engagement, we also distill the Pin's
OmniSage~\cite{omnisage} embedding $\mathbf{y}$, a graph embedding that combines
content features with user engagement, into the quantized representation, while keeping $\mathbf{x}$
as the only input so that SIDs do not change when OmniSage is updated.
Training details are in Appendix~\ref{sec:appendix-rqvae}.

\begin{figure*}[t]
  \centering
  \includegraphics[width=\textwidth]{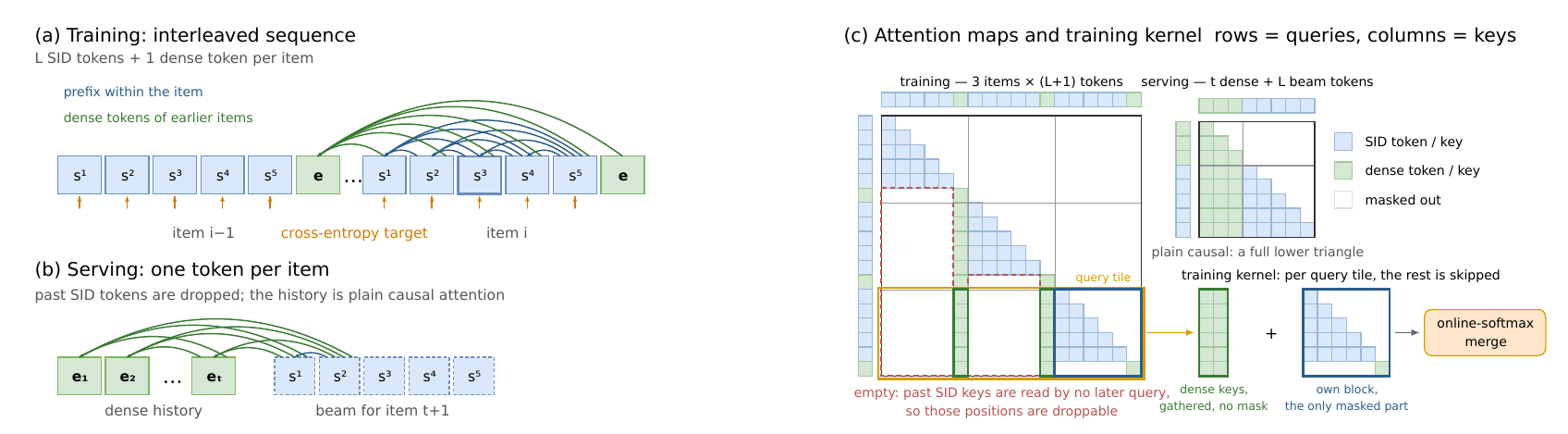}
  \caption{\textbf{(a)} Interleaved sequence in training. \textbf{(b)}
  Sequence at serving, with past SID tokens dropped. \textbf{(c)} Training and
  serving attention maps, and the training kernel's split of a query tile
  (Section~\ref{sec:mask-kernel}).}
  \label{fig:mask}
\end{figure*}

\subsection{Interleaved sequence modeling}
\label{sec:mask}

We use dense tokens on the input side, where one token per item keeps a long
history within the context budget, and SIDs on the output side, which remove
the ANN index and let decoding shape diversity (Section~\ref{sec:diversity}).
An encoder--decoder would place the two in separate backbones, but its decoder
then sees the history only through cross-attention and receives one next-item
label per forward pass. We therefore keep both representations within a single
causal backbone, with the following designs.

\subsubsection{Interleaved layout}
During training the user sequence is represented by dense embeddings and SID
tokens interleaved.
Each history item $i$ contributes $L{+}1$ consecutive tokens (Figure~\ref{fig:mask}(a)): its SID tokens
$s^{1}_{i},\ldots,s^{L}_{i}$, embedded by the per-level codebooks, followed by
one dense token $e_{i}$.
All $L{+}1$ tokens share the position index $i$, so the positional encoding
counts items rather than tokens.

\subsubsection{Asymmetric mask}
Attention is restricted by a single predicate, stated as three rules
(Figure~\ref{fig:mask}(c)).
\begin{enumerate}[leftmargin=1.4em,itemsep=1pt,topsep=2pt]
  \item \textbf{The dense stream is closed.} A dense query $e_{i}$ attends only
    to dense keys $e_{j}$ with $j\le i$, never to an SID token.
  \item \textbf{SID queries read dense history plus their own prefix.} The
    level-$\ell$ SID token of item $i$ attends to $s^{1}_{i},\ldots,s^{\ell}_{i}$
    and to every $e_{j}$ with $j<i$.
  \item \textbf{Nothing else.} An SID token never attends to another item's SID
    tokens, nor to the dense token of its own item.
\end{enumerate}
Rule~3's second half is a causality constraint imported from serving: at decode
time the next item's own embedding is precisely the unknown, so a training-time
path from $s^{\ell}_{i}$ to $e_{i}$ would be unreproducible at inference.
Rules~1 and~3 together mean that no position outside item $i$ ever attends to
item $i$'s SID tokens.
Those tokens can therefore be dropped from the sequence without altering any
dense hidden state, and hence without altering the distribution over the next
identifier.
So training runs at $L{+}1$ tokens per item and serving at one, over the
same context and computing the same function of the history: at decode time only
the dense tokens are kept, and the sole SID tokens present are those of the beam
being extended (Figure~\ref{fig:mask}(b)).

\subsubsection{Supervision}
The pretraining objective of the sequence model is SID cross-entropy, applied
per level with logits weight-tied to the corresponding RQ-VAE codebook and
uniform weights across levels.
Every SID token is a prediction target; dense positions carry a reserved level
identifier and are never targets, so the model is never asked to regress an
embedding.
A sequence of $S$ tokens, or $|H|=S/(L{+}1)$ items, therefore yields $|H|\,L$ next-token
predictions per forward pass, where an encoder--decoder supervises only the
codes of a single target item.
Each prediction is conditioned on the dense history and the true shallower
codes of its own item, so level $\ell$ is trained given $H$ and
$c_{1},\ldots,c_{\ell-1}$, the same condition under which beam search generates
it at inference.

\subsubsection{Sparsity-aware attention kernels}
\label{sec:mask-kernel}
The asymmetric mask makes serving cheap, but training still runs at $L{+}1$ tokens per item, and at long contexts attention
dominates the step time. Most of this attention is masked. In
Figure~\ref{fig:mask}(c), the unmasked entries are the dense-token columns,
which every later token reads, and a small triangle per item for its SID
tokens. Off-the-shelf sparse attention cannot exploit this structure.  FlexAttention~\cite{flexattention} skips a block only if it is fully
masked. Every block below the diagonal contains part of a dense column, so no
block is skipped and the cost equals that of full causal attention.

We instead wrote custom forward and backward kernels for this mask. Each query tile
contains whole items, so all of its rows attend to the same earlier dense
tokens. The kernel gathers these tokens and computes their attention without a
mask. It then computes attention among the tile's own tokens, where each SID
token reads the shallower codes of its own item; this is the only part that
needs the mask. Online
softmax~\cite{dao2023flashattention2fasterattentionbetter} combines the two. The
backward pass splits the work by key: dense keys reduce to causal attention
over the dense tokens, and each SID key is read only by tokens of its own item.
Compared with TIGER, the attention cost is lower by a factor of about
$L^{2}/(L{+}1)$ despite the extra token per item
(Appendix~\ref{sec:appendix-kernel}). At serving, the mask reduces to standard
causal attention, so these kernels are not needed.

We further drop the level-$L$ SID token from the training input, since the
dense token that follows it is never a prediction target
(Figure~\ref{fig:arch}) and no later token attends to it. This keeps every target and
shortens the sequence to $L$ tokens per item. Together, the two changes raise
training throughput by $94\%$ over a plain TIGER model at the same
context length.

\subsection{Feature embeddings and goal conditioning}
\label{sec:features}

\begin{figure*}[t]
  \centering
  \includegraphics[width=\textwidth]{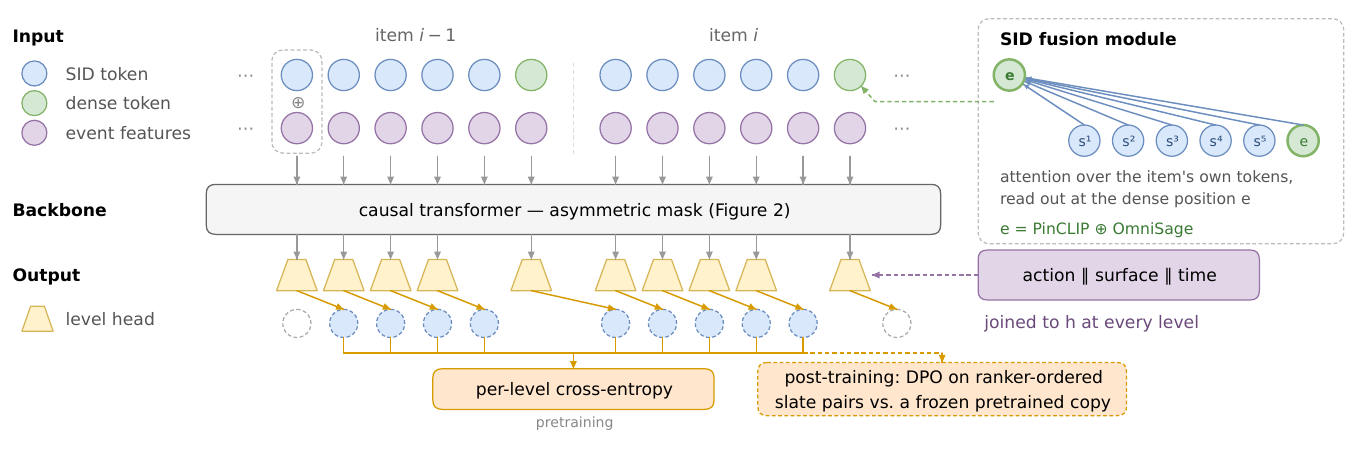}
  \caption{The interleaved sequence model: event features at every position,
  SID fusion producing the dense token, goal conditioning at every level head,
  per-level cross-entropy, and DPO post-training
  (dashed; Section~\ref{sec:post-training}).}
  \label{fig:arch}
\end{figure*}

We adopt the feature set of PinRec~\cite{pinrec} and
UniPinRec~\cite{unipinrec}, and the outcome-conditioned generation of PinRec.
Under the interleaved layout the features are attached to each of an item's $L$
SID tokens and to its dense token, and the conditioning is applied once per SID
token.
Figure~\ref{fig:arch} shows the resulting dataflow.

\paragraph{Item features.}
By Rule~1 the dense token is the only channel through which item $i$ reaches
later positions, so it is where the item's representation is assembled.
The item representation $\mathbf{e}_{i}$ concatenates two pretrained
embeddings, each projected and $\ell_2$-normalized: a content-only PinCLIP
multimodal embedding~\cite{pinclip}, and a graph embedding OmniSage~\cite{omnisage}
that incorporates collaborative engagement signal.
Because the graph embedding never reaches the tokenizer, collaborative signal
can be carried in the sequence and updated as engagement evolves, with no
refresh of the SIDs.

\paragraph{SID fusion module.}
The item's SID embeddings serve twice: they occupy their own positions in the
sequence, and they are also folded into $\mathbf{e}_{i}$ by a \emph{SID fusion
module} --- a single causal attention layer over the item's $L{+}1$ tokens
$[s^{1}_{i},\ldots,s^{L}_{i},\mathbf{e}_{i}]$, read out at the dense position.
Fusion adds no content beyond $\mathbf{e}_{i}$, since the tokenizer quantizes
the same multimodal embedding; what it adds is a \emph{discrete} view of the
item --- a position in the corpus hierarchy, over which the model can
generalize when generating coarse prefixes, and a stable identity that frequent
items can be memorized by.
The feature therefore costs no extra input positions.

\paragraph{Event features.}
Each position also carries the timestamp, surface and action type of its event.
These are concatenated with the position's own embedding and projected, so
position $p$ of item $i$ receives
\begin{equation}
  \label{eq:dense-token}
  z_{p} = \mathrm{Proj}\bigl[\, f_{p} \;\Vert\;
    \psi(t_{i},t_{i-1}) \;\Vert\;
    \mathbf{u}_{\mathrm{surface}} \;\Vert\;
    \mathbf{u}_{\mathrm{action}} \,\bigr],
\end{equation}
where $f_{p}$ is the codebook embedding $\mathbf{q}^{(\ell)}_{c_{\ell}}$ at the
level-$\ell$ SID position and $\mathbf{e}_{i}$ at the dense position. Following
PinRec~\cite{pinrec}, $\psi$ encodes the absolute timestamp and the gap to the
preceding position with sinusoids (Appendix~\ref{sec:appendix-timestamp}).

\paragraph{Goal conditioning.}
Like PinRec, we generate toward a desired outcome rather than an arbitrary
next item. The goal vector concatenates embeddings of the target action and
surface with an encoding of the request time,
$\mathbf{g} = [\mathbf{u}_{\mathrm{action}} \Vert \mathbf{u}_{\mathrm{surface}}
\Vert \psi(t,t_{\mathrm{prev}})]$, and is appended to the hidden state $h$ before each level head,
\begin{equation}
  \label{eq:cond-head}
  p_\theta(c_\ell \mid c_{1:\ell-1}, H, \mathbf{g})
    \;\propto\;
    \exp\bigl(\mathbf{E}^{(\ell)} \, o_\ell\bigl([\,h \,\Vert\, \mathbf{g}\,]\bigr)\bigr),
\end{equation}
where $o_\ell$ is the level-$\ell$ head and $\mathbf{E}^{(\ell)}$ its output
embeddings. Conditioning every level lets the goal steer both the coarse
levels, which pick a region of the corpus, and the fine levels within it. At
serving, the caller supplies the action, surface and time it wants, in the
same form as the event features of past interactions. Conditioning controls
what kind of item is generated, and the diversity penalty
(Section~\ref{sec:diversity}) spreads the beam across items of that kind.

\subsection{Decode-time steering: prefix diversity}
\label{sec:diversity}

\begin{figure*}[t]
  \centering
  \includegraphics[width=\textwidth]{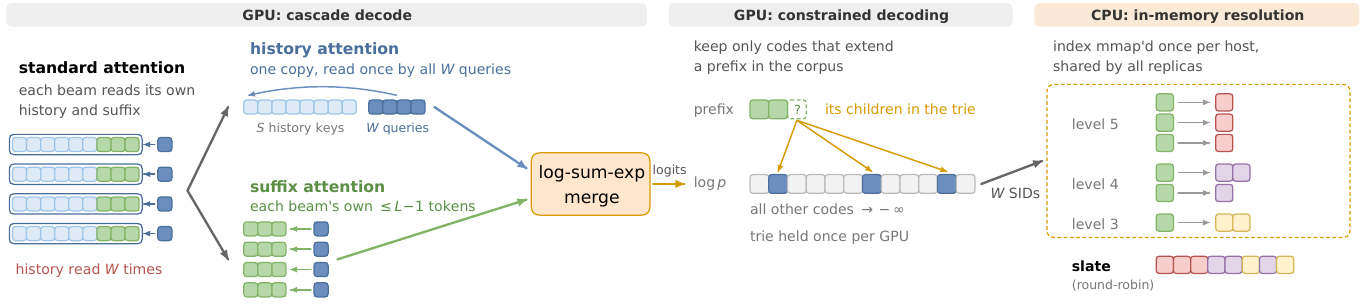}
  \caption{The serving path: cascade decode, which splits beam attention into a
  shared history pass and a per-beam suffix pass; constrained decoding over the
  SID trie; and in-memory resolution of SIDs to items.}
  \label{fig:cascade}
\end{figure*}

\paragraph{Prefix collapse.}
The beam score accumulates over levels, so at each level the ranking is
dominated by the prefixes already chosen. The $W$ best continuations of one
strong prefix usually outscore the best continuation of any other, and top-$W$
selection keeps the beam inside a few subtrees. Because prefixes are semantic
(Section~\ref{sec:semantic-ids}), the retained beams are also similar in
content: at $W{=}100$ they span only $11.9$ distinct level-$1$ codes
(Table~\ref{tab:div-sweep}).

\paragraph{Why not temperature sampling.}
Temperature sampling does not recover the lost spread: both spread and
lookup@$L$ degrade monotonically in $T$ (Table~\ref{tab:div-sweep}). Sampling
perturbs selection but not scoring, so beams admitted on improbable codes are
outscored at the next level and their slots return to the dominant subtrees.
The final level has no later level to prune a poor draw, so that is where
temperature hallucinates and validity collapses: the generated identifiers name no corpus item. We therefore keep decoding greedy and introduce a prefix diversity penalty into the selection
rule.

\paragraph{Hierarchical prefix penalty.}
At level $\ell$, each parent beam $w$ and code $v$ form a candidate with score
$s(w,v)=S_{w}+\log p_\theta(v\mid c^{(w)}_{1:\ell-1},H,\mathbf{g})$, where
$S_{w}$ is the parent's accumulated score. Instead of taking the top $W$
candidates at once, we select the $W$ beams one at a time. Let $n_{w}$ be the
number of beams already taken from parent $w$, $v^{(j)}_{w}$ its $j$-th best
code, and $m(w,w')$ the length of the longest common prefix of parents $w$ and
$w'$. The next beam is
\begin{equation}
  \label{eq:diversity}
  \arg\max_{w}\ \Bigl\{\,
    s\bigl(w,\,v^{(n_{w}+1)}_{w}\bigr)
    \;-\; \lambda \sum_{w'} m(w,w')\, n_{w'}
  \,\Bigr\}.
\end{equation}
Each parent is penalized by the beams already chosen that share its prefix,
weighted by the length of the shared prefix, so beams that agree at deeper
levels are penalized more. With $\lambda{=}0$ the rule reduces to top-$W$
selection. The penalties are updated incrementally, with $W$ additions per
selected beam, and each level runs as one fused GPU kernel
(Appendix~\ref{sec:appendix-diversity}).
Section~\ref{sec:div-results} evaluates the trade-off between spread and
recall.

\subsection{Post-training: training-time steering}
\label{sec:post-training}
\label{sec:grpo}

The pretrained model of Equation~\ref{eq:model} is trained with a next-token
loss on engaged items only, which has two problems. First, it has no
contrastive loss to push the model away from negatives. Second, it is
teacher-forced on individual codes, with no item-level objective that matches
autoregressive decoding at inference. As a result, what the model retrieves is
poorly aligned with what the ranker scores highly. We therefore post-train with
DPO~\cite{rafailov2023dpo}.

Preference pairs are sampled from the serving logs, where every impression is
stored with the utility the production ranker computed at request time. Within
each slate we sort items by this utility and pair the top three against the
bottom three. Whereas OneRec~\cite{onerec} trains a separate reward model to
label such pairs, we use the ranker's logged scores directly, so no reward model
is trained or served.

Let $s_\theta(c_{1:\ell}) = \sum_{t\le\ell} \log p_\theta(c_t \mid c_{1:t-1}, H)$
be the log-probability the policy assigns to the prefix $c_{1:\ell}$ given
the history $H$, and
$s_{\mathrm{ref}}$ the same under the frozen pretrained model. For a pair
$(c^{+}, c^{-})$ the DPO loss at level $\ell$ is
\begin{equation}
  \mathcal{L}^{\ell}_{\mathrm{DPO}} = -\log\sigma\!\Big(\beta\big[
    \big(s_\theta(c^{+}_{1:\ell}) - s_\theta(c^{-}_{1:\ell})\big) -
    \big(s_{\mathrm{ref}}(c^{+}_{1:\ell}) - s_{\mathrm{ref}}(c^{-}_{1:\ell})\big)
  \big]\Big),
  \label{eq:dpo}
\end{equation}
and we sum it over $\ell = 1,\dots,L$ so that the coarse levels are trained
as well. Subtracting the
reference margin means the loss stops pushing on pairs the pretrained model
already orders correctly. The total loss is
$w\sum_\ell \mathcal{L}^{\ell}_{\mathrm{DPO}} + \mathcal{L}_{\mathrm{NTP}}$,
where $\mathcal{L}_{\mathrm{NTP}}$ is the pretraining next-identifier loss on
the engaged items of the same slates. The anchor is necessary: with the DPO
term alone, the policy drifts from the reference and recall collapses within a
few thousand steps.

\section{Serving Infrastructure}\label{sec:infra}

The serving stack inherits from the PinRec~\cite{pinrec} and
UniPinRec~\cite{unipinrec} deployments: a Triton inference server driving the
transformer forward pass, with their feature fetch, request batching and ensemble routing
unchanged. SID generation and Pin retrieval add requirements that stack does
not meet, which we address with three changes. \emph{Cascade decode} applies
shared-prefix attention~\cite{hydragen} to beam decoding, so a request walks
its history once rather than once per beam (Section~\ref{sec:paged-cache}).
\emph{Constrained decoding} restricts every level to codes the corpus
contains, using the vectorized static trie of STATIC~\cite{su2026vectorizing} held
once per host (Section~\ref{sec:constrained}).
\emph{Resolution} maps each generated identifier to items through an
in-memory, host-local index (Section~\ref{sec:local-index}).
Section~\ref{sec:serving-numbers} reports the three composed.

\subsection{Cascade decode}
\label{sec:paged-cache}

During decoding, all beams of a request share the user history, about $10^3$ tokens at serving,
and differ only in the at most $L{-}1$ SID tokens generated so far. A contiguous
KV cache gives each beam its own copy of the history, so memory grows as
$B \times W \times S$. PagedAttention~\cite{pagedattention} shares the history in
fixed-size blocks, but two costs remain. A beam still copies every block it
writes to, and since $L$ is far smaller than the block size (a multiple of
$256$ in the flash-attention kernel we use), most of each copied block is
wasted. And each of the $B \times W$ beams is a separate query row that reads the
whole history again, so the cost grows linearly in $W$.

\begin{figure}[t]
  \centering
  \includegraphics[width=\linewidth]{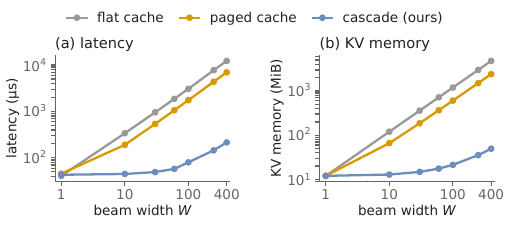}
  \caption{(a) Latency and (b) KV memory of one attention call at one layer,
  on one A10G ($B{=}4$, $S{=}1018$).}
  \label{fig:cascade-perf}
\end{figure}

We instead fold the beam into the query dimension. The query becomes a
$(B, W, H, d_h)$ tensor attending to a history indexed by $B$ alone, so each
request reads its history once for all $W$ beams, with no causal mask since
the beams are siblings at the same decode step. Each beam then attends to its
own suffix buffer, which holds only the current item's generated tokens. The
two attentions return their outputs and log-sum-exps, which combine exactly
into attention over the full sequence~\cite{hydragen}; our kernel fuses the
suffix attention and this merge into a single launch
(Figure~\ref{fig:cascade}).

Figure~\ref{fig:cascade-perf} compares one attention call against a flat
per-beam cache and a paged cache. The three are nearly identical at $W{=}1$.
As the beam widens, flat and paged attention grow linearly in $W$, whereas
cascade grows only $5\times$ from $W{=}1$ to $W{=}400$; it is
$4.25\times$ faster than paged attention at $W{=}10$ and
$32\times$ faster at $W{=}400$. Memory follows the same pattern: cascade
holds $12.6\times$ less KV than the paged cache at $W{=}30$ and
$48.7\times$ less at $W{=}400$, since only the small suffix buffers grow
with the beam.

\subsection{Constrained decoding}
\label{sec:constrained}

Unconstrained beam search can emit any of the $K$ codes at every level, so a
generated SID may name no item in the corpus. Without a constraint, only about
$66\%$ of generated SIDs resolve at level $5$ ($87\%$ at level $4$ and $98\%$
at level $3$), and the misses fall back to a coarser prefix than the one the
model chose. Checking validity on the host at every level would stall the GPU
inside the decode step.

Following STATIC~\cite{su2026vectorizing}, we keep the trie of valid prefixes on
the GPU as static tensors. The first two levels are dense validity masks of
size $K$ and $K{\times}K$. Deeper levels are sparse, so we store them in
compressed sparse row form: prefixes are sorted, an offsets array marks where
each prefix's children start, and a values array lists the children. A beam's
state is the index of its prefix, so one batched gather reads the allowed codes
of all beams, and after a beam picks a code, its new state is the start of its
run plus the rank of the code within it. Decoding thus needs no pointer chasing
and no host round trip, and runs inside the compiled decode graph. Codes
outside the allowed set are masked to $-\infty$ after the log-softmax, leaving
top-$k$ expansion and the diversity penalty (Section~\ref{sec:diversity})
unchanged, and every beam then resolves at level $5$. All replicas on a GPU
share one copy of the trie through CUDA IPC.
The constraint adds 0.30 and 0.59\,ms to decoding at $W{=}100$ and
$W{=}400$ ($9.8\%$ and $5.9\%$), nearly independent of corpus size,
and the trie occupies only 168 and 1{,}368\,MiB of GPU memory for
corpora of 10\,M and 100\,M items (Table~\ref{tab:trie} in Appendix~\ref{sec:appendix}).

\subsection{In-memory serving}
\label{sec:local-index}

Once decoded, each beam's identifier and its coarser prefixes must be resolved
to items. We serve this from a host-resident index, rebuilt daily. As in
Section~\ref{sec:constrained}, each level is stored in compressed sparse row
form: the keys are the SID prefixes in sorted order, the offsets mark where
each prefix's item signatures start and end, and the values are the signatures
themselves, so a lookup is one vectorized binary search plus a slice. Storing
item signatures in addition to transitions requires roughly
$10\times$ the decoding trie's memory. The indices are written once and
memory-mapped read-only, so a single physical copy per host serves every
replica.

Resolving at the finest level may not yield enough items, since a single
level-$L$ prefix usually holds only a few. A beam that comes up short falls
back to its coarser prefixes, widening the neighborhood one level at a time.
Candidates are interleaved across beams round-robin, so a crowded leaf cannot
monopolize the slate.

\subsection{End to end: the composed system}
\label{sec:serving-numbers}

We compare PinGER at five beam widths with PinRec at the same user sequence
length and matched per-host load (Table~\ref{tab:e2e}). The two share a serving
stack and differ only in model inference and Pin resolution. PinGER is faster at
every beam width, mostly because retrieval moves onto the host: the host-local
index replaces PinRec's remote nearest-neighbor call, which costs
31\,ms at p99, and stays under 20\,ms at p99 at every width.

\begin{table}[t]
  \caption{Serving latency (ms) and GPU memory against PinRec at
  ${\sim}$200 QPS per host, on an AWS g6e host (NVIDIA L40S).}
  \label{tab:e2e}
  \small
  \setlength{\tabcolsep}{4pt}
  \begin{tabular}{lccccc}
    \toprule
    & e2e p50 & e2e p99 & model p99 & retrieval p99 & mem.\ \% \\
    \midrule
    embedding & 79 & 110 & 24 & 31 & 40 \\
    \midrule
    $W{=}30$ & 48 & 73 & 12 & 4 & 52 \\
    $W{=}60$ & 51 & 77 & 13 & 8 & 52 \\
    $W{=}100$ & 54 & 81 & 14 & 12 & 52 \\
    $W{=}250$ & 61 & 92 & 20 & 19 & 52 \\
    $W{=}400^\dagger$ & 60 & 93 & 23 & 16 & 53 \\
    \bottomrule
  \end{tabular}
  \\[2pt]
  \footnotesize Model: embedding forward pass or beam decoding; retrieval:
  remote nearest-neighbor call or host-local resolution.
  $^\dagger$Diversity penalty disabled.
\end{table}

Widening the beam adds latency to both decode and resolution, since each
additional beam is another prefix to resolve, but not memory: GPU memory stays
flat across beam widths, the serving counterpart of
Figure~\ref{fig:cascade-perf}(b), because cascade attention holds the history
once however wide the beam.

\section{Experiments}\label{sec:experiments}
\subsection{Shared setup}
\label{sec:setup}

\paragraph{Datasets.}
We use the session-centric data of UniPinRec~\cite{unipinrec}. Each example
pairs a user's historical action sequence, the Pins they engaged with across
surfaces together with the action type, surface and time of each engagement,
with a session from the target surface whose engaged Pins serve as targets.
Examples are sampled by session rather than by user, so that the data reflects
actual traffic. Evaluation uses held-out users in a later, non-overlapping time
window, disjoint from training, and generated identifiers are resolved against
the full production corpus.

\paragraph{Metrics.}
Offline we report Recall@$k$ for $k \in \{10, 100\}$ at beam width $W{=}100$:
the fraction of held-out next interactions whose SID is among the top $k$
generated identifiers, micro-averaged over users. For the diversity analysis we
also report distinct@$1$, the number of distinct level-$1$ prefixes among the
$W$ beams, and lookup@$L$, the share of generated SIDs that resolve to items in
the production corpus. Online we report production A/B metrics
(Section~\ref{sec:online}).

\paragraph{Implementation details.}
The backbone is a decoder-only transformer with $4$ layers, hidden size $768$
and $12$ attention heads of dimension $64$, with as many key--value heads as
query heads. The SID fusion module is a single attention layer with $4$ heads.
Each SID has $L{=}5$ levels, each drawn from a codebook of $K{=}2048$ codes.
Evaluation decodes greedily unless stated otherwise.

\subsection{Offline results}
\label{sec:offline}

\subsubsection{Main comparison}
\label{sec:main-results}
\label{sec:seq-length}

Table~\ref{tab:main} compares PinGER with PinRec and TIGER at matched cost.
Context length is counted in tokens, and the two SID layouts spend it differently:
TIGER uses $L{=}5$ tokens per history item and ours $L{+}1$, so at a fixed
training context length TIGER sees slightly more items. At serving the asymmetry
reverses: TIGER still reads every SID token, while PinGER reads one dense token
per item, six times fewer. At the same 1{,}024 serving tokens PinGER's
Recall@100 is therefore $9.5\%$ higher than TIGER's and
$3.9\times$ PinRec's, and to match TIGER's recall it needs only
170 tokens.

This efficiency matters because history length matters. Both TIGER and PinGER
gain about
$10\%$ in Recall@100 as the training context length grows from
1{,}024 to 6{,}144 tokens, and neither has saturated: each step up
in context length still improves both (Table~\ref{tab:sweep}). With its
efficient token usage, PinGER can capture these gains from longer history
without a serving cost that scales with the number of SID tokens per item.

\begin{table}[t]
  \caption{Recall@100 of PinGER, TIGER and PinRec at the same serving cost and
  at the same recall, beam width $100$.}
  \label{tab:main}
  \small
  \setlength{\tabcolsep}{3.5pt}
  \begin{tabular}{llrrcl}
    \toprule
    & model & context length & serving & R@100 (\%) & \\
    \midrule
    \multicolumn{6}{l}{\emph{Same serving cost}} \\
    & PinRec          & 1{,}024 & 1{,}024          & 1.01          & \\
    & TIGER           & 1{,}024 & 1{,}024          & 3.65          & \\
    & \textbf{PinGER} & 6{,}144 & 1{,}024          & \textbf{4.00} & \textbf{$+9.5\%$} \\
    \midrule
    \multicolumn{6}{l}{\emph{Same recall}} \\
    & TIGER           & 1{,}024 & 1{,}024          & 3.65          & \\
    & \textbf{PinGER} & 1{,}024 & \textbf{170}     & 3.64          & \textbf{$6.0\times$ fewer} \\
    \bottomrule
  \end{tabular}
  \\[2pt]
  \footnotesize Context length is the training context length; serving is the
  number of tokens read at serving time.
\end{table}

\subsubsection{Ablations}
\label{sec:ablation}

Table~\ref{tab:ablation} in Appendix~\ref{sec:appendix} removes one component of Section~\ref{sec:features} at
a time, and we report the change in Recall@100. SID fusion matters most:
without it Recall@100 falls by $27.0\%$, since the dense token is then
built from the item features alone, without the item's own SID tokens. Goal
conditioning is next, at $17.0\%$. The two item embeddings matter far
less. Removing PinCLIP leaves recall essentially unchanged ($\Delta{\approx}0$):
PinCLIP is redundant once its SIDs are in the input, since the SIDs are
quantized from the PinCLIP embedding (Section~\ref{sec:semantic-ids}) and SID
fusion carries them into the dense token. Removing OmniSage costs $1.2\%$, as its engagement
signal is only partly distilled into the SIDs.

\subsubsection{Diversity penalty}
\label{sec:div-results}

Table~\ref{tab:div-sweep} sweeps the prefix-diversity penalty $\lambda$ on the
$6{,}144$-token model and, for comparison, the sampling temperature $T$. The
penalty raises distinct level-$1$ prefixes among the
$100$ beams from 11.9 under greedy decoding to 93.6 at
$\lambda{=}3$, while the top-10 lookup rate barely moves and recall falls only
modestly. Greedy decoding concentrates the beam in the most probable subtrees,
where the single held-out item usually lies, so spreading it trades next-item
recall for coverage of the user's other interests. Temperature, by contrast,
fails on both counts, for the reason given in Section~\ref{sec:diversity}. It
unexpectedly \emph{lowers} diversity, to 3.5 distinct prefixes at
$T{=}1.5$, and only $13\%$ of the top-10 identifiers exist in the
corpus, against $74\%$ under greedy decoding, so most of what
temperature generates cannot be served. Recall@100 falls
accordingly, by $43\%$ at $T{=}0.7$ and $85\%$ at $T{=}1.5$,
against $28.5\%$ for the mildest penalty. The penalty therefore
outperforms temperature sampling: at comparable settings it yields more diverse and
more valid beams, and retains more recall.

\begin{table}[t]
  \caption{Diversity penalty $\lambda$ versus sampling temperature $T$, beam
  width $100$. Recall and lookup in \%.}
  \label{tab:div-sweep}
  \small
  \setlength{\tabcolsep}{2.5pt}
  \begin{tabular}{lccccc}
    \toprule
    decoding & R@10 & R@100 & distinct@1 & lookup@$L$ top10 & lookup@$L$ all \\
    \midrule
    greedy          & 1.54 & 4.00 & 11.9 & 74.3 & 60.6 \\
    \midrule
    $\lambda{=}0.3$ & 1.35 & 2.86 & 39.4 & 74.1 & 61.6 \\
    $\lambda{=}0.5$ & 1.26 & 2.47 & 49.9 & 73.9 & 59.6 \\
    $\lambda{=}1.0$ & 1.14 & 2.07 & 68.3 & 73.4 & 55.3 \\
    $\lambda{=}3.0$ & 0.93 & 1.42 & 93.6 & 72.3 & 50.5 \\
    \midrule
    $T{=}0.7$       & 0.74 & 2.27 & 7.0  & 50.3 & 36.5 \\
    $T{=}1.5$       & 0.20 & 0.60 & 3.5  & 13.2 & 8.0  \\
    \bottomrule
  \end{tabular}
\end{table}

\subsubsection{Post-training}
\label{sec:post-training-results}
We also post-train the same model with the DPO of
Section~\ref{sec:post-training}. Offline,
post-training lifts recall by $2.3\%$ and the difference in the
log-likelihood between engaged and impression-only items by $86\%$.
Online, the retrieved identifiers align better with the ranker, with a
$10\%$ improvement in candidate survival rate in the final slate.

\subsection{Online A/B testing}
\label{sec:online}

We deploy PinGER on two surfaces, Related Pins and notifications, each in an
A/B experiment against PinRec~\cite{pinrec}, the production embedding
retriever, at a fixed candidate budget. All reported
lifts are statistically significant at the 5\% level (two-sided test).

\subsubsection{Related Pins}
\label{sec:ab-rp}

Related Pins recommends Pins related to the one a user is viewing. We replace
PinRec's nearest-neighbor retrieval path with PinGER,
decoding with beam width $60$ and diversity penalty $\lambda{=}3.0$, and
resolving identifiers from level $5$ down to level $3$, over a two-week experiment. Table~\ref{tab:ab} shows gains on
the surface that carry over site-wide: surface saves rise by $0.62\%$
and grid clicks by $0.94\%$, and site-wide saves by $0.61\%$. The
largest lift is on fresh content, where saves of Pins created in the last
$28$ days rise by $3.60\%$. We attribute this to the identifiers being
derived from content embeddings (Section~\ref{sec:semantic-ids}): a new Pin is
retrievable as soon as it has an SID, whereas PinRec depends
on engagement embeddings that take weeks to stabilize. Removing the
nearest-neighbor index also retires its serving fleet, saving its
infrastructure cost, and lowers Related Pins latency by $0.7\%$.

\paragraph{Effect of the diversity penalty.}
Users are also shown, and adopt, a wider range of
interests, with $0.56\%$ more novel interests impressed and
$0.43\%$ more interest domains adopted, the online counterpart of the beam diversity in
Section~\ref{sec:div-results}. These gains depend on the penalty's strength. A separate
arm with a weaker penalty, $\lambda{=}1.0$, lifts novel interests by only
$0.36\%$, interest domains by $0.10\%$ and grid clicks by
$0.77\%$, and leaves surface impressions ($+0.37\%$) and saves
($-0.38\%$) without a significant change. The stronger penalty costs
more recall offline (Table~\ref{tab:div-sweep}), but online it is the setting
that pays.

\begin{table}[t]
  \caption{Online A/B results on Related Pins and notifications, relative to
  PinRec at an equal candidate budget.}
  \label{tab:ab}
  \small
  \setlength{\tabcolsep}{2pt}
  \begin{tabular}{@{}lllc@{}}
    \toprule
    surface & type & metric & lift \\
    \midrule
    Related Pins  & Surface    & saves / impressions                & $+0.62$ / $+0.70\%$ \\
                  &            & grid clicks                           & $+0.94\%$ \\
                  &            & time spent                         & $+0.34\%$ \\
                  &            & fulfilled sessions                 & $+0.35\%$ \\
    \cmidrule{2-4}
                  & Site-wide  & saves / impressions                & $+0.61$ / $+0.59\%$ \\
                  &            & sessions $>$5 min                   & $+0.22\%$ \\
    \cmidrule{2-4}
                  & Freshness  & saves / impr., Pins $<$28 days     & $+3.60$ / $+4.90\%$ \\
    \cmidrule{2-4}
                  & Diversity  & novel interests / domains          & $+0.56$ / $+0.43\%$ \\
    \cmidrule{2-4}
                  & System     & latency                            & $-0.7\%$ \\
    \midrule
    Notifications & Engagement & email click / push open rate       & $+0.90$ / $+0.25\%$ \\
                  &            & saves from notifications          & $+0.62\%$ \\
    \bottomrule
  \end{tabular}
\end{table}

\paragraph{Segment analysis.}
An in-depth analysis traces the engagement gains to three groups
(Table~\ref{tab:segments} in Appendix~\ref{sec:appendix}).
(1)~\textbf{High-utility Pins}: among the top decile of Pins by
engagement, PinGER's candidates take a larger share and have a
$5.4\%$ higher save rate, even though the ranker places them lower
than PinRec's.
(2)~\textbf{Undervalued Pins}: PinGER surfaces more Pins with little prior
engagement, and these Pins have a $13\%$ higher save rate. This
breaks the feedback loop common in traditional recommenders, where items
that are already engaged are surfaced ever more often.
(3)~\textbf{Most active users}: the users who engage most receive a larger
share of PinGER's candidates and save them $12\%$ more often, which
we attribute to a more diverse and fresher feed.

\subsubsection{Notifications}
\label{sec:ab-notif}

Pinterest's email and push notifications recommend Pins to bring users back to
the platform. Here PinGER supplies half of the candidate pool, replacing
PinRec candidates so that the total budget is unchanged, decoding with beam width $100$ and diversity penalty
$\lambda{=}1.0$. Table~\ref{tab:ab} shows consistent gains in
notification engagement. Unlike Related Pins, a stronger penalty, $\lambda{=}3.0$, gave a smaller
click-rate gain, $0.76\%$, and raised unsubscribes, so we deploy the
milder setting: beyond a point, diversity in a short notification slate costs
more than it returns.

\section{Conclusion}\label{sec:conclusion}

We presented PinGER, a generative retriever deployed at Pinterest. An
asymmetric attention mask over interleaved SID and dense tokens lets the model
train with SID supervision at every position and serve the history at one token
per item, and a serving stack built for wide beams removes nearest-neighbor
search from the request path.

Future work includes sharing a single SID-based backbone between retrieval and
ranking, in the spirit of UniPinRec~\cite{unipinrec}, and applying the
interleaved layout to LLM-based recommenders~\cite{plum,hllm}, whose
multi-token item representations limit the history that fits in their context.

\bibliographystyle{ACM-Reference-Format}
\bibliography{base}

\appendix
\section{Additional detail}\label{sec:appendix}

\begin{table}[!h]
  \caption{Full context-length sweep, beam width $100$: Recall@100
  of PinRec, and of TIGER and PinGER each trained at three context lengths,
  with the serving length each requires.}
  \label{tab:sweep}
  \small
  \setlength{\tabcolsep}{4pt}
  \begin{tabular}{rlrcr}
    \toprule
    context length & model & items & R@100 (\%) & \textbf{serving} \\
    \midrule
    1{,}024 & PinRec & 1{,}024 & 1.01 & 1{,}024          \\
    \midrule
    1{,}024 & TIGER  & 204     & 3.65 & 1{,}024          \\
            & PinGER & 170     & 3.64 & \textbf{170}     \\
    \midrule
    2{,}048 & TIGER  & 409     & 3.88 & 2{,}048          \\
            & PinGER & 341     & 3.78 & \textbf{341}     \\
    \midrule
    6{,}144 & TIGER  & 1{,}228 & 4.06 & 6{,}144          \\
            & PinGER & 1{,}024 & 4.00 & \textbf{1{,}024} \\
    \bottomrule
  \end{tabular}
  \\[2pt]
  \footnotesize Context length and serving in tokens; items is the history
  length that context length holds.
\end{table}

\begin{table}[!h]
  \caption{Cost of constrained decoding on one A10G. Latency ms,
  resident trie MiB; uniform-random corpora bound both columns from above.}
  \label{tab:trie}
  \small
  \setlength{\tabcolsep}{4pt}
  \begin{tabular}{lcccc}
    \toprule
    & $W{=}100$ & $W{=}400$ & resident & B/SID \\
    \midrule
    unconstrained        & 3.007    & 10.115   & ---              & ---         \\
    \midrule
    10\,M corpus  & $+0.295$ & $+0.596$ & 168.0     & 17.6 \\
    100\,M corpus & $+0.298$ & $+0.593$ & 1{,}367.8 & 14.3 \\
    \bottomrule
  \end{tabular}
\end{table}

\begin{table}[!h]
  \caption{Segment analysis on Related Pins: PinGER's candidates against
  PinRec's. Share is the relative change in the segment's share of the
  retriever's impressions; save rate is the relative change in the segment's
  save rate.}
  \label{tab:segments}
  \small
  \setlength{\tabcolsep}{4pt}
  \begin{tabular}{@{}lcc@{}}
    \toprule
    segment & share & save rate \\
    \midrule
    top 10\% of utility        & $+23\%$ & $+5.4\%$ \\
    \quad of these, in the top 5 slots & $-19\%$ & $+4.2\%$ \\
    Pins with the fewest prior saves$^{*}$ & $+20\%$ & $+13\%$ \\
    most active users$^{\dagger}$         & $+46\%$ & $+12\%$ \\
    \bottomrule
  \end{tabular}
  \\[2pt]
  \footnotesize $^{*}$Lowest decile of saves over the prior seven days.
  $^{\dagger}$Top segment by saves over the prior seven days.
\end{table}

\begin{table}[!h]
  \caption{Ablations at a $6{,}144$-token context, beam width $100$.
  Recall in \%; $\Delta$ is relative to the full model.}
  \label{tab:ablation}
  \small
  \setlength{\tabcolsep}{4pt}
  \begin{tabular}{lcccc}
    \toprule
    & R@10 & $\Delta$ & R@100 & $\Delta$ \\
    \midrule
    full model         & 1.54 & ---                  & 4.00 & ---                  \\
    $-$ PinCLIP        & 1.56 & $\approx0^{*}$       & 4.04 & $\approx0^{*}$       \\
    $-$ OmniSage       & 1.50 & $-2.4\%$      & 3.95 & $-1.2\%$      \\
    $-$ goal conditioning & 1.38 & $-10.7\%$ & 3.32 & $-17.0\%$     \\
    $-$ SID fusion     & 0.98 & $-36.6\%$     & 2.92 & $-27.0\%$     \\
    \bottomrule
  \end{tabular}
  \\[2pt]
  \footnotesize $^{*}$No measurable change ($+1.6\%$ and $+1.1\%$).
\end{table}

\subsection{Tokenizer training}
\label{sec:appendix-rqvae}
The MLP encoder maps $\mathbf{x}$ to a latent $\mathbf{z}_{e}$.
A residual quantizer then produces the discrete code $c$ and a quantized vector $\mathbf{z}_{q}=\sum_{\ell=1}^{L}\mathbf{q}^{(\ell)}_{c_{\ell}}$, from which the decoder reconstructs $\hat{\mathbf{x}}$.
Let $\mathbf{r}_{\ell}$ be the residual at level $\ell$ and $\mathbf{q}_{\ell}$ the selected code vector.
We train with the standard RQ-VAE objective
\begin{equation}
  \mathcal{L}_{\mathrm{RQ}}
  =
  \underbrace{\lVert \mathbf{x}-\hat{\mathbf{x}}\rVert_{2}^{2}}_{\text{reconstruction}}
  +
  \sum_{\ell=1}^{L}
  \Bigl(
    \underbrace{\lVert \mathrm{sg}[\mathbf{r}_{\ell}]-\mathbf{q}_{\ell}\rVert_{2}^{2}}_{\text{codebook}}
    +
    \beta\,
    \underbrace{\lVert \mathbf{r}_{\ell}-\mathrm{sg}[\mathbf{q}_{\ell}]\rVert_{2}^{2}}_{\text{commitment}}
  \Bigr),
\end{equation}
where $\mathrm{sg}[\cdot]$ denotes the stop-gradient and $\beta$ weights the commitment term.
Codebooks are updated by gradient descent on this loss.

\paragraph{Engagement distillation.}
User interactions induce Pin--Pin similarities that $\mathbf{x}$ does not capture.
Let $\mathbf{y}$ denote the Pin's OmniSage~\cite{omnisage} embedding, a graph embedding learned from user engagement.
Feeding $\mathbf{y}$ into the RQ-VAE would make SIDs move whenever OmniSage is updated, so we keep $\mathbf{x}$ as the only input and distill engagement into $\mathbf{z}_{q}$ via $\mathbf{y}$ instead.
A distillation head produces $\mathbf{y}_{q}=h_{\psi}(\mathbf{z}_{q})$.
In a batch of size $B$ we form temperature-scaled cosine similarities, mask the diagonal, and match neighborhood distributions with a KL divergence:
\begin{equation}
  \mathcal{L}_{\mathrm{eng}}
  =
  \frac{1}{B}
  \sum_{i=1}^{B}
  D_{\mathrm{KL}}
  \bigl(
    p^{\tau}(\mathbf{y}_{i}\,\Vert\,\{\mathbf{y}_{j}\}_{j\neq i})
    \,\big\Vert\,
    p^{\tau}(\mathbf{y}_{q,i}\,\Vert\,\{\mathbf{y}_{q,j}\}_{j\neq i})
  \bigr).
\end{equation}
This term must flow into the codebooks, so we do not use residual $k$-means, which has no gradient path from $\mathcal{L}_{\mathrm{eng}}$ into the codes.

\paragraph{Progressive masking and dead codes.}
Following PLUM~\cite{plum}, we sample a cutoff $k\sim\mathrm{Unif}\{1,\ldots,L\}$ at training time and form $\mathbf{z}_{q}$ from only the first $k$ residual levels, so earlier codebooks are forced to compress more of the content and engagement signal.
To keep the $K$-way vocabularies from collapsing we track an EMA of assignment counts and reinitialize any code whose EMA falls below one from randomly permuted encoder latents in the current batch.

\subsection{Attention cost of the training kernels}
\label{sec:appendix-kernel}
For $|H|$ items, the kernels of Section~\ref{sec:mask-kernel} compute about
$(L{+}1)|H|^{2}/2$ attention scores. TIGER computes about $L^{2}|H|^{2}/2$,
since each token attends to all codes of the earlier items, and FlexAttention
computes about $(L{+}1)^{2}|H|^{2}/2$ under our mask. Compared with TIGER, the
attention cost is therefore lower by a factor of $L^{2}/(L{+}1)$ despite the
extra token per item.

\subsection{Timestamp gaps in the interleaved layout}
\label{sec:appendix-timestamp}
The gap encoded by $\psi$ in Equation~\ref{eq:dense-token} needs care in the
interleaved layout. At the first SID position of an item it is the true gap
$t_{i}-t_{i-1}$, but at the other positions the predecessor belongs to the same
item and the gap is zero. The dense token instead copies the encoding of the
first SID position, which matches serving, where the dense token directly
follows the previous item's dense token.

\subsection{Computing the prefix penalty}
\label{sec:appendix-diversity}
The rule of Equation~\ref{eq:diversity} is cheap to compute. A parent's $k$-th selected candidate is always
its $k$-th best code, so each parent's codes are sorted once per level and
only the counts $n_{w}$ change. Selecting a winner $w^{\star}$ increments only
$n_{w^{\star}}$, and since the penalty is linear in $n$, the penalty vector is
updated by adding the column $m(\cdot,w^{\star})$, which takes $W$ additions
per step and never stores the $W{\times}W$ matrix. The $W$ steps depend on each
other and must run in sequence, so the loop for each level runs as one fused
GPU kernel.

\end{document}